\documentclass[conference]{IEEEtran}
\IEEEoverridecommandlockouts
\usepackage{cite}
\usepackage{amsmath,amssymb,amsfonts}
\usepackage{algorithm}
\usepackage{algpseudocode}
\usepackage{graphicx}
\usepackage{textcomp}
\usepackage{xcolor}
\def\BibTeX{{\rm B\kern-.05em{\sc i\kern-.025em b}\kern-.08em
    T\kern-.1667em\lower.7ex\hbox{E}\kern-.125emX}}
\begin{document}

\title{An Efficient Algorithm for Computing Mountain Prominence in Almost Linear Time}

\author{\IEEEauthorblockN{George Alex Dumitrescu}
\IEEEauthorblockA{\textit{Faculty of Computer Science} \\
\textit{Alexandru Ioan Cuza University of Iasi}\\
Iasi, Romania \\
george.dumitrescu@student.uaic.ro}
\and
\IEEEauthorblockN{Paul Flavian Diac}
\IEEEauthorblockA{\textit{Faculty of Computer Science} \\
\textit{Alexandru Ioan Cuza University of Iasi}\\
Iasi, Romania \\
paul.diac@uaic.ro}
}

\IEEEoverridecommandlockouts 
\IEEEpubid{
\makebox[\columnwidth]{979-8-3503-6813- 0/24/\$31.00~\copyright2024 IEEE\hfill} \hspace{\columnsep}\makebox[\columnwidth]{ }
}

\maketitle

\begin{abstract}
Prominence is one of the most important measurements in topography and mountaineering. This paper describes an efficient, almost linear time algorithm for computing mountain prominence for all peaks on Earth using digital elevation models (DEMs). It builds on top of a classic algorithm and leverages the observation that only a few peaks have their prominence determined by a relatively distant other mountain. Thus, the classic algorithm can be adapted to memorize and use less information without the loss of correctness. The algorithm is demonstrated using 3 arcsecond real-life data from SRTM datasets. Its importance is underscored by the increasing accuracy of Earth mapping methods and the corresponding growth in the amount of data that must be processed to compute prominence.
\end{abstract}

\begin{IEEEkeywords}
big data, algorithms, mountain prominence, digital elevation model (DEM), SRTM data
\end{IEEEkeywords}

\section{Introduction}
Prominence, along with Elevation and Isolation, is among the most commonly used and important metrics in mountain classification and mountaineering. While elevation denotes the height of a point on the Earth's surface, isolation and prominence are significant because they measure how mountains relate to each other and how a mountain dominates its surroundings.

Prominence, also known as shoulder drop or relative height, is usually defined in two equivalent ways:
\begin{itemize}
\item The vertical distance from a peak summit to the lowest contour line encircling it without containing a higher peak;
\item The minimum vertical distance one must descend from a summit to reach a higher point.

\end{itemize}

\begin{figure}[htbp]
\centering
\includegraphics[width=0.48\textwidth]{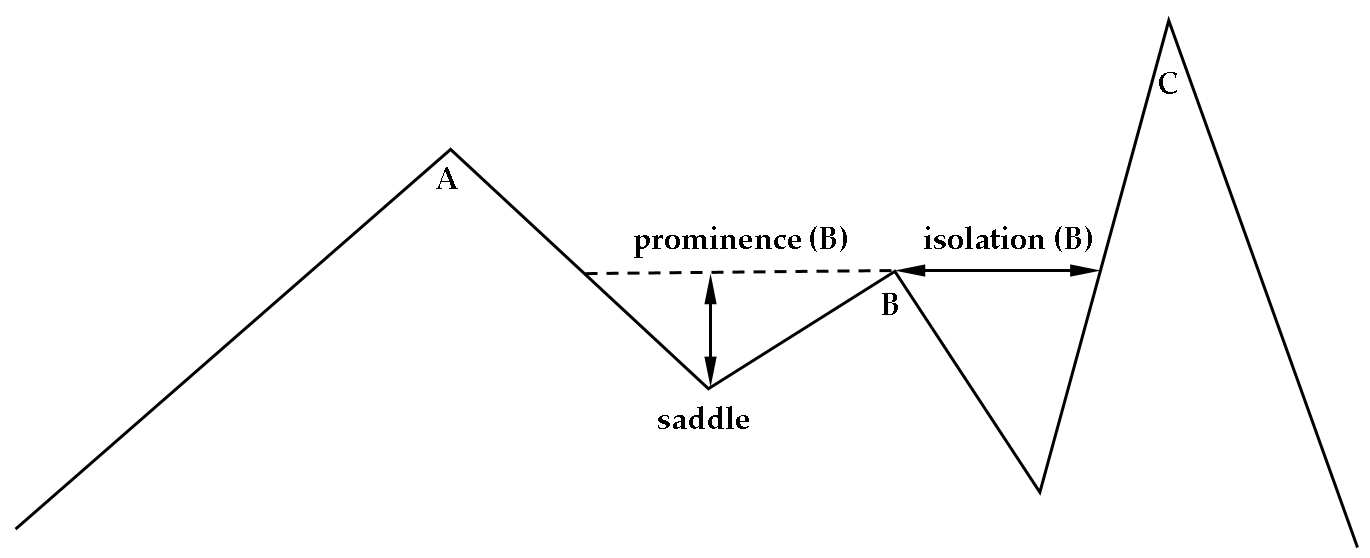}
\caption{Prominence and isolation visualized}
\label{fig1}
\end{figure}

While some algorithms have been developed to compute prominence directly from topographic maps, this paper uses the second definition as its foundation. When considering all possible paths connecting a peak to any higher terrain, prominence is determined by identifying the path with the highest lowest point. This point is known as \textit{key col} or \textit{saddle}. The prominence is the difference in elevation between the summit and the saddle. Prominence, isolation and the saddle point are all ilustrated in Fig.~\ref{fig1}.

Researchers and mountaineers are greatly interested in a mountain's prominence because it is a key indicator of the summit's significance. Many use prominence as a criterion for including peaks in mountain peak lists \cite{b1}. Moreover, prominence serves as a valuable indicator of the subjective importance of a summit. For instance, Viștea Mare Peak, the third-highest summit in Romania, illustrates the significance of prominence. Due to its proximity to Moldoveanu Peak, the highest mountain in Romania, connected by a short ridge, many do not consider it a standalone climb.

Digital Elevation Models (DEMs) are essential tools for surface analysis \cite{b2}. Within the scope of this paper, they represent the Earth's surface as a 2D matrix where each cell is defined by latitude and longitude lines, with its value indicating the corresponding elevation on Earth. The size of these cells determines the resolution of the DEM.

As the resolution of DEMs increases, more efficient algorithms are needed to compute prominence. Data with a resolution of 3 arcseconds, equivalent to approximately 90 meters at the equator, is already available for the entire Earth. In some areas, data with a resolution of 1 arcsecond or even 0.3 arcseconds is accessible.

\section{Previous and related work}

Topographic prominence has been used since the 1930s to classify and discover peaks \cite{b3}. It was measured by survey or by examining topographic maps. With these "by hand" methods, most of the known prominence values have been calculated and used to create mountain lists.

In the early 2000s, the first computer programs that automatically compute mountain prominence were created. A pioneer in this area was Edward Earl, who introduced innovative approaches and developed Winprom \cite{b4}, which could compute prominence over larger areas and extract meaningful information about the structure of mountain ranges.

More recently, Kirmse and de Ferranti took the ideas left behind by Edward Earl and implemented an algorithm capable of calculating the prominence and isolation of every mountain in the world \cite{b5}. 

The core of their implementation is creating a tree-like data structure starting from a DEM, where the nodes are peaks and the saddles connecting the nodes are edges. This approach gives a very good characterization of the topographic surface, the tree taking the shape of the underlying mountain range. The speed of the algorithm is given by its multithreading capabilities and a complicated pruning method that eliminates from the computation peaks that have the prominence under a certain threshold. Using this algorithm on DEM data, they compared the results with already-known prominence values and found significant differences in certain instances.

Unfortunately, algorithms encounter limitations due to data voids and inaccuracies. For example, the SRTM data used in this paper has a height error of more than 16 m for 10\% of the data \cite{b6}. However, progress is being made in enhancing data coverage and precision. Thus, new innovative algorithms are needed to accommodate this growth in the quantity of data.

Throughout the world, communities of "Peak Baggers" are passionate about mountains and their prominence values. Consequently, multiple lists of mountains with prominence data are available on the internet. Among these, the database from peakbagger.com stands out as one of the most important and widely accepted sources \cite{b7}.

\section{Water sweep algorithm}

\subsection{High-Level Description}
This subsection introduces the base algorithm, which will be referred to in this paper as "water sweep," a name given by the authors. The water sweep algorithm that represents the starting point of this paper is straightforward and intuitive. One can imagine the surface of interest to be covered in imaginary water that gradually recedes, forming "islands" as it does so. When two islands merge, it signifies the highest elevation for which a path exists between the peaks included in the two islands. The merging point is a saddle.

\begin{figure}[htbp]
\centering
\includegraphics[width=0.38\textwidth]{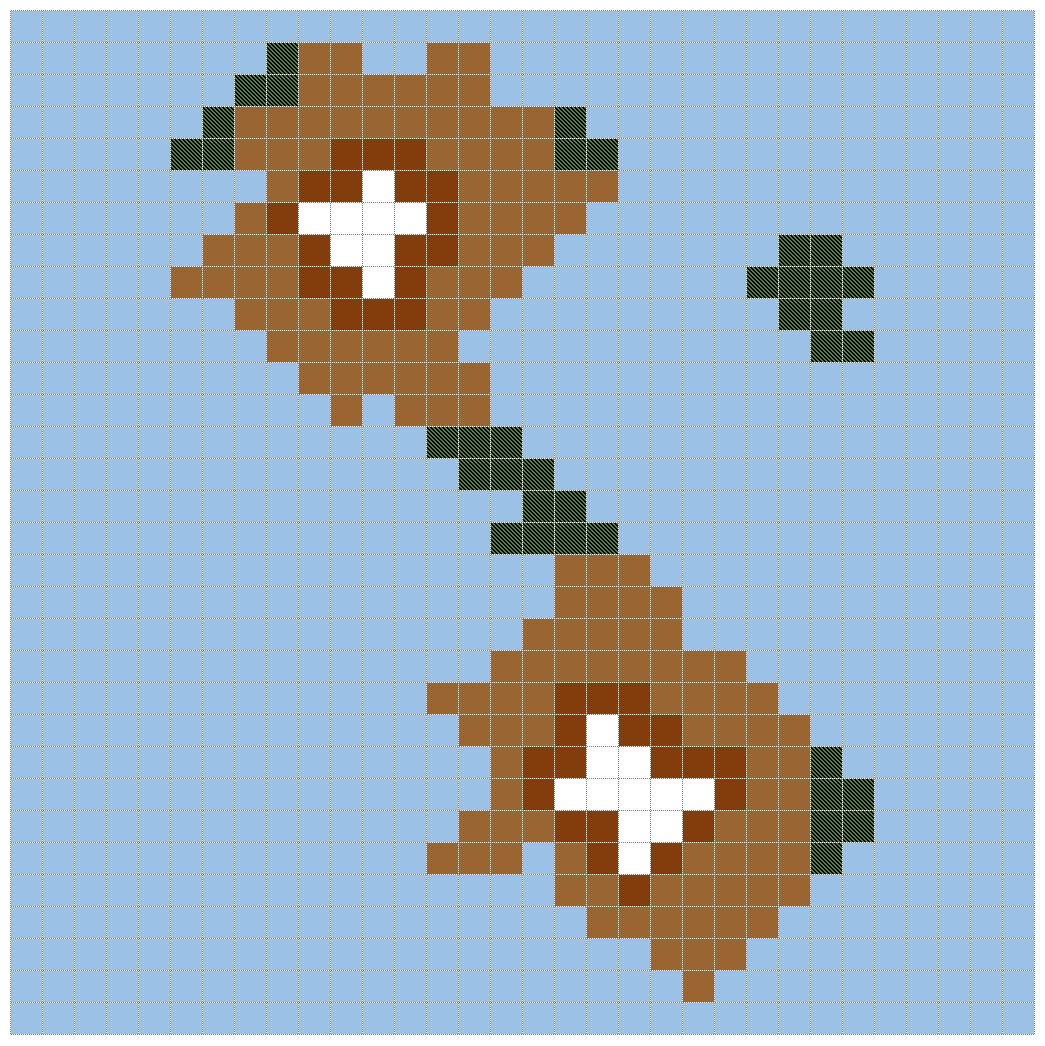}
\caption{A step of the water sweep algorithm. Blue - water, Brown - existing islands, Green - newly added areas, White - dominant peaks}
\label{fig2}
\end{figure}

In this iterative algorithm, at a moment where two islands connect, the highest peaks from each island are of interest. The highest peak from an island will be called a \textit{dominant peak}. The saddle found at the moment when the two islands merge will determine the prominence of the smaller of the two dominant peaks. One iteration of the algorithm is exemplified in Fig.~\ref{fig2}.

The highest peak across the entire surface will have, by definition, a prominence equal to its height. This, in the context of Earth, means that Everest will have 8,848m prominence. 

\subsection{Implementation}

A Digital Elevation Model (DEM) can be viewed as a 2D array with elevation values on which lowering the water level is simulated.

The first step of the algorithm is to divide the matrix into areas of cells of the same height called \textit{plateaus}. This can be done by defining when two cells are neighbors and using either Depth-First Search (DFS) or Breadth-First Search (BFS) algorithms \cite{b8}. In the authors' implementation, BFS is utilized, and two cells are considered neighbors if they share an edge or a corner (8 neighbors for each cell). Each area is treated as a node in a graph, and adjacency between areas determines connections in the graph. Islands generated by the algorithm represent subgraphs within this graph.

The second step of the algorithm is to identify peaks. A peak is considered to be an area of connected cells that is higher than all areas immediately adjacent to it. For these areas, prominence will be calculated.

The third step is sorting the areas by height. Most DEMs, including the one used in this paper, use meters or feet as units of measure, and the elevation is rounded to the nearest integer. Given Earth's maximum elevation of 8,848 meters, an efficient sorting algorithm suited to this context is counting sort.

The fourth and final step of the algorithm is to simulate the water level lowering. Initially, the algorithm starts with an empty graph. When the water level goes below the height of an area, the associated node is added alongside all its edges to nodes that have already been added to the graph. For each connected component, the dominant peak is memorized, and if the added node connects two different connected components, the prominence of one of the peaks can be computed. Naturally, an area can connect multiple different connected components. This can be implemented using the disjoint-set data structure, also known as union-find.

\subsection{Complexity}

The complexity of the algorithm is almost linear. If the matrix is considered to have $n$ cells, the complexity of dividing the matrix in plateaus is $\mathcal{O}(n)$ and sorting is $\mathcal{O}(n + H)$, where $H$ is the maximum height on the surface. The time of performing find operations or union operations $\mathcal{O}(n\alpha(n))$ where $\alpha$(n) is the inverse Ackermann function \cite{b9}. There are $\mathcal{O}(n)$ find and union operations because there are, at most, $n$ nodes, and each node is added once.

Therefore, the overall complexity is $\mathcal{O}(n\alpha(n) + H)$. Thus, the algorithm is nearly optimal, with the complexity almost linear.

\subsection{Problems regarding data}

For this paper, 3 arcsecond SRTM data collected by Jonathan de Ferranti \cite{b10} was used. This data is organized in tiles of 1$^{\circ}$x1$^{\circ}$ degrees latitude. Each tile is a matrix of size 1,201 x 1,201, resulting in a total of 1,226,221 samples per tile. On Earth, there are 360$^{\circ}$ of longitude and 180$^{\circ}$ of latitude, with each degree corresponding to one tile. This results in a total of 64,800 tiles. Thus, the total number of values that must be stored and processed for the entire Earth is 93,467,584,800. Such a quantity of data greatly exceeds the storage and processing capacity of a typical program running on an ordinary computer. In the next section, an algorithm that addresses this issue will be presented. The main idea of the algorithm is to treat each tile individually, find the information that has an impact on other tiles, and then use this data to merge different tiles. 

\section{Applying the water sweep algorithm inside tiles}

\subsection{Peak classification}

If the water sweep algorithm, described in Section II, is applied to a single tile, each peak from within the tile will have a prominence value. However, these prominence values can be incorrect. A peak can have its prominence determined by a higher peak from another tile, and thus, the actual prominence value is different from the one inside found within the tile. Thus, the question of determining if the prominence value found within the tile is the actual value arises.

Assume that at some point in the water sweep algorithm, two islands that do not touch the edge of the tile are merged. The prominence value is correct for the smaller dominant peak from the two islands. This is because to reach the edge of the tile, a descent lower than the height of the saddle must be made. Every path from the dominant peak to another higher area outside the tile will have the lowest point lower than the found saddle and will not be significant for the final prominence.

If both islands touch the edge of the tile, the true prominence of the dominant peaks remains unknown. Both peaks can have the prominence determined by a higher peak outside the tile. Even if a value for the prominence for the smaller dominant peak is calculated by the water sweep algorithm, it is only an upper bound for the true prominence.

If only one of the islands touches the edge of the tile, two cases can be identified. For one of them, the prominence value is correct; for the other, it is not. The cases are as follows:

\begin{enumerate}
    \item The dominant peak from the island that does not touch the edge is lower. In this case, the value found by the water sweep algorithm is the correct one. This is true because the height of the saddle is equal to the highest lowest point of any path from the lower dominant peak to the outside of the tile. 
    \item The dominant peak from the island that touches the edge is lower. In this case, similar to when two islands that touch the edge connect, the prominence value that is found for the lower dominant peak is just an upper bound for the true prominence.
\end{enumerate}

\begin{figure}[htbp]
\centering
\includegraphics[width=0.48\textwidth]{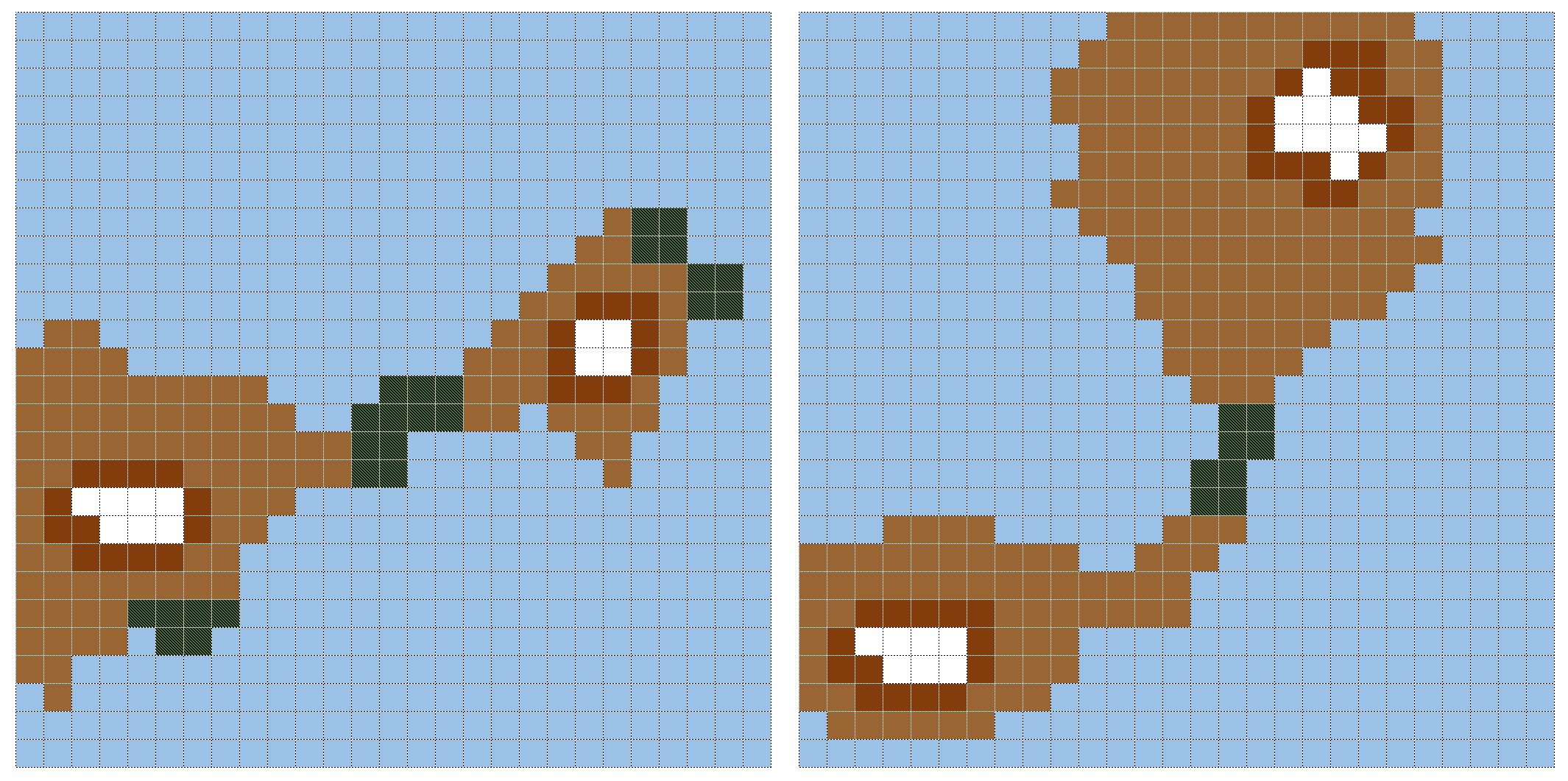}
\caption{Examples for the two cases: On the left, only one of the two islands touches an edge; on the right, both islands touch the edge}
\label{fig3}
\end{figure}

The case where only one of the two islands that are merged touches the edge and the case where both islands that are merged touch the edge is shown in Fig.~\ref{fig3}.

By analyzing these cases, two types of peaks can be identified:
\begin{itemize}
    \item \textit{Inside peaks}: peaks for which the water sweep algorithm finds the correct prominence value without taking into account the neighboring tiles;
    \item \textit{Outside peaks}: peaks for which a provisional prominence value is known and for which  other tiles need to be taken into account in order to verify and update the prominence value if that is the case.
\end{itemize}

\subsection{Outside peaks. Proof of sufficiency}

The number of outside peaks for a tile tends to be relatively low due to the natural formation of mountain ranges. Through experimentation done for this paper using the water sweep algorithm across various tiles worldwide, employing different resolutions of the DEM, it was observed that for a tile containing $n$ x $n$, the number of outside peaks is lower than $n$, with more than 97\% of peaks from within a tile being inside peaks.

\begin{figure}[htbp]
\centering
\includegraphics[width=0.4\textwidth]{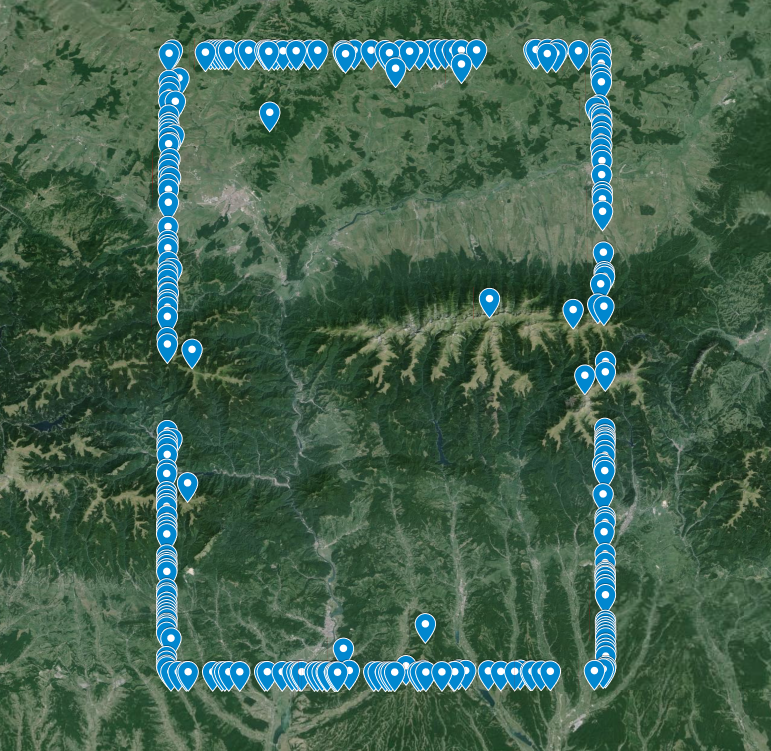}
\caption{Outside peaks for tile N45E024}
\label{fig4}
\end{figure}

The following example utilizes tile N45E024 from the SRTM dataset analyzed in this paper. This tile covers the area 45-46$^{\circ}$N and 24-25$^{\circ}$E and contains the highest peak in Romania, Moldoveanu. On this tile, there are 1,193,347 areas (plateaus) and 10,119 peaks, but only 264 outside peaks, of which 201 are exactly on the edge of the tile. This example is shown in Fig.~\ref{fig4}, where outside peaks are highlighted.

An important observation that is the foundation for the algorithm described in this paper is the fact that for an outside peak, the higher peak that determines its prominence can only be another outside peak. Thus, to compute the true prominence values for the outside peaks, inside peaks can be ignored, significantly lowering the amount of data that must be stored and processed.

This can be proven by contradiction in the following way. Assume there exists an inside peak A, which is the peak that determines the prominence of a peak outside its tile (peak B). However, the fact that peak A is an inside peak implies that there was a moment in the original algorithm where the island where it was dominant did not touch any of the edges and was united with another island with a higher dominant peak. This means that in the moment when the island that contains the inside peak A touches any of the edges, another peak is the dominant peak of that island, which will be taken into consideration as the peak that determines the prominence for the outside peak B.

\subsection{Information that must be saved}

The previous subsection proves that only the outside peaks need to be stored and used in computing the prominence of each other. This means that in the context of the SRTM dataset used in this paper and the observation described in the previous subsection, at most, 77,824,800 outside peaks must be stored for the whole Earth, which results in a great reduction in size. In practice, when applying the algorithm, the number of outside peaks is much lower, and a method for eliminating outside peaks that become inside peaks when merging tiles is described in the following section.

\begin{figure}[htbp]
\centering
\includegraphics[width=0.3\textwidth]{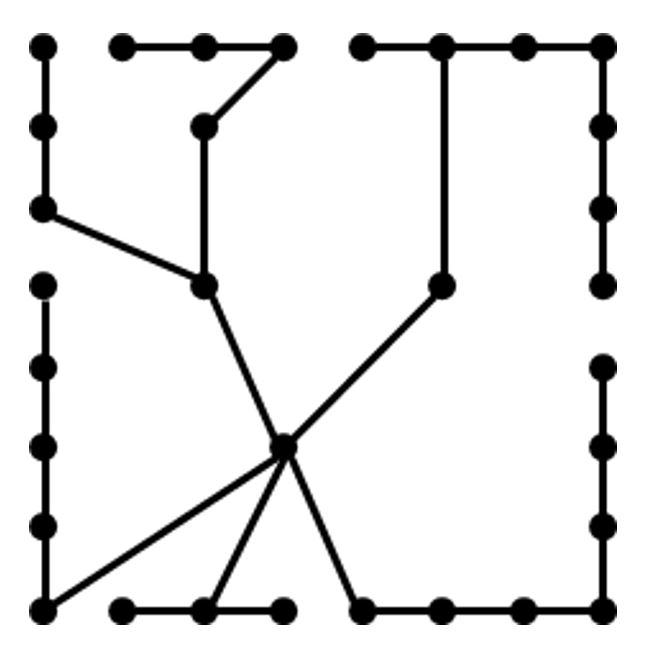}
\caption{Example of tree generated by the algorithm}
\label{fig5}
\end{figure}

The main objective is to merge the information from neighboring tiles. To do this, outside peaks and edge areas are stored in memory in a weighted tree data structure. Nodes in the tree are all the areas that are on the edge of the tile and all outside peaks. The edges of the tree connect dominant peaks and are added when two islands that touch the edge of the tile are united or when an island that does not touch the edge of the tile is united with an island that touches the edge of the tile but with a lower dominant peak (the case where outside peaks appear). The weight of the edge is equal to the height of the saddle that unites the two islands. Fig.~\ref{fig5}. is an example of what such a tree could look like. All the areas on the edges and all the outside peaks from a tile are nodes.

By this construction, the tree has the property that the unique path between two vertices corresponds to the path with the highest lowest point between the two underlying areas. The height of this point is the minimum value of the edges encountered in the path. This property is used when merging tiles. 

\section{Merging tiles}

\subsection{Algorithm for neighboring tiles}

SRTM tiles are typically processed to be seamless, meaning that they repeat values along all edges. Therefore, for example, the last column of a tile should be equal and represent the same area as the first column of the neighboring tile. This is only sometimes the case for the dataset used for this paper. Thus, the following preprocessing of the data was necessary. For a tile, the last column is replaced with the first column of the next tile with the same latitude. The same is applied to the last line of the tile and the first line of the next tile with the same longitude. 

The algorithm for merging two tiles uses a combination of all the properties and observations previously described in the paper. 

Assume that two neighboring tiles that share an edge are merged. Areas on the common edge will represent the connection between the two trees. The path of minimum drop between two peaks from different tiles will be found in this new graph, which is obtained by merging the trees. The path will pass through one or more of the shared areas.

The water sweep algorithm can be applied to this graph. This time, it is applied by sorting the edges of the graph and adding them in decreasing order of height. Similar to when the water sweep algorithm was applied on a single tile, connected components are formed, and when two islands are merged, the prominence of the lower dominant peak can be calculated. As in the case of single tiles, islands can contain nodes that are on the edge. In this case, the edge for the two merged tiles is formed from the union of all previous edges except the one between them, which is eliminated in this merge. Therefore, outside peaks can still remain outside peaks if the path to the higher peak passes through nodes on the edge. However, some nodes, for which the path to the peak that determines the prominence does not pass through edge nodes, become inside nodes, and their correct prominence is calculated.

The same algorithm described in Subsection IV.C is used to create a tree data structure that has the same priorities and stores information about the nodes in the merge. This tree contains the peaks that remain outside peaks and the areas that are on the edges of the merge.

This tree, which results after the merging operation, can be used further in the next steps of the algorithm. Groups of tiles for which this tree is known can be merged by transforming the common areas in nodes and applying the algorithm described above. These merges are done until there remains a single group of tiles that contains all the tiles from the surface of interest.

\subsection{Divide and Conquer Optimization and final complexity}

The land surface that is processed by the proposed algorithm is considered to be a rectangular area defined by four coordinates: $minLat$, $maxLat$, $minLon$, and $maxLon$. Any irregular shape of tiles can be filled to fit a rectangle by adding "empty" tiles.

\begin{algorithm}
\caption{Prominence Finder algorithm}
\small 
\begin{algorithmic}[1]
\Function{DQ}{$minLat, maxLat, minLon, maxLon$}
    \If{$(minLat = maxLat$ \textbf{and} $minLon = maxLon)$}
        \State \Call{readTile}{$minLat, minLon$}
        \State \Call{applyWaterSweepAlgorithm}{ }
        \State $tree \gets$ \Call{buildTree}{ }
        \State \Return $tree$
    \EndIf
    
    \State $mLat \gets (minLat + maxLat) / 2$
    \State $mLon \gets (minLon + maxLon) / 2$

    \If{$(minLat \neq maxLat$ \textbf{and} $minLon \neq maxLon)$}
        \State $uL \gets$ \Call{DQ}{$minLat, mLat, minLon, mLon$}
        \State $uR \gets$ \Call{DQ}{$minLat, mLat, mLon + 1, maxLon$}
        \State $up \gets$ \Call{mergeTrees}{$uL, uR$}

        \State $dL \gets$ \Call{DQ}{$mLat + 1, maxLat, minLon, mLon$}
        \State $dR \gets$ \Call{DQ}{$mLat + 1, maxLat, mLon + 1, maxLon$}
        \State $down \gets$ \Call{mergeTrees}{$dL, dR$}

        \State \Return \Call{mergeTrees}{$up, down$}
    \EndIf

    \If{$(minLat = maxLat)$}
        \State $L \gets$ \Call{DQ}{$minLat, maxLat, minLon, mLon$}
        \State $R \gets$ \Call{DQ}{$minLat, maxLat, mLon + 1, maxLon$}
        \State \Return \Call{mergeTrees}{$left, right$}
    \EndIf

    \State $up \gets$ \Call{DQ}{$minLat, mLat, minLon, maxLon$}
    \State $down \gets$ \Call{DQ}{$mLat + 1, maxLat, minLon, maxLon$}
    \State \Return \Call{mergeTrees}{$up, down$}
\EndFunction
\end{algorithmic}
\end{algorithm}

In a tile group merge operation, the water sweep algorithm is used, which consists of sorting the nodes and using a union-find data structure. The complexity of merging two groups of tiles would be $\mathcal{O}(H + N \alpha(N))$, where $N$ is the number of nodes in the graph. Therefore, it is imperative to choose TileTrees merges efficiently. This can be achieved by using the Divide and Conquer technique. At each step, the current surface is divided into four equal subsurfaces (when possible), and then the results from these subsurfaces are merged. This process is repeated recursively until reaching the base case of a single tile. For the base case, the water sweep algorithm is applied to extract the tree of outside peaks. The number of merges done by the algorithm is $\mathcal{O}(T)$, where $T$ is the number of tiles in the analyzed surface. This procedure is shown in Algorithm 1.

If a tile is considered to be of size $n$ x $n$, and $T$ is the number of the number of tiles, the final complexity of the merging algorithm would be $\mathcal{O}(T\log(T)H+ \log(T) \cdot (T \cdot n))$. This is because every node is used in $\mathcal{O}(log(T))$, and there are $\mathcal{O}(T \cdot n)$ nodes in the trees generated by the water sweep algorithm. The complexity is not substantial compared to the complexity of applying the water sweep algorithm on each tile $\mathcal{O}(T \cdot (n \cdot n + H))$; thus, the execution time of the algorithm would greatly benefit from the use of multithreading.

\section{Experimental results on a large surface area}

\subsection{Data used}

\begin{figure}[htbp]
\centering
\includegraphics[width=0.47\textwidth]{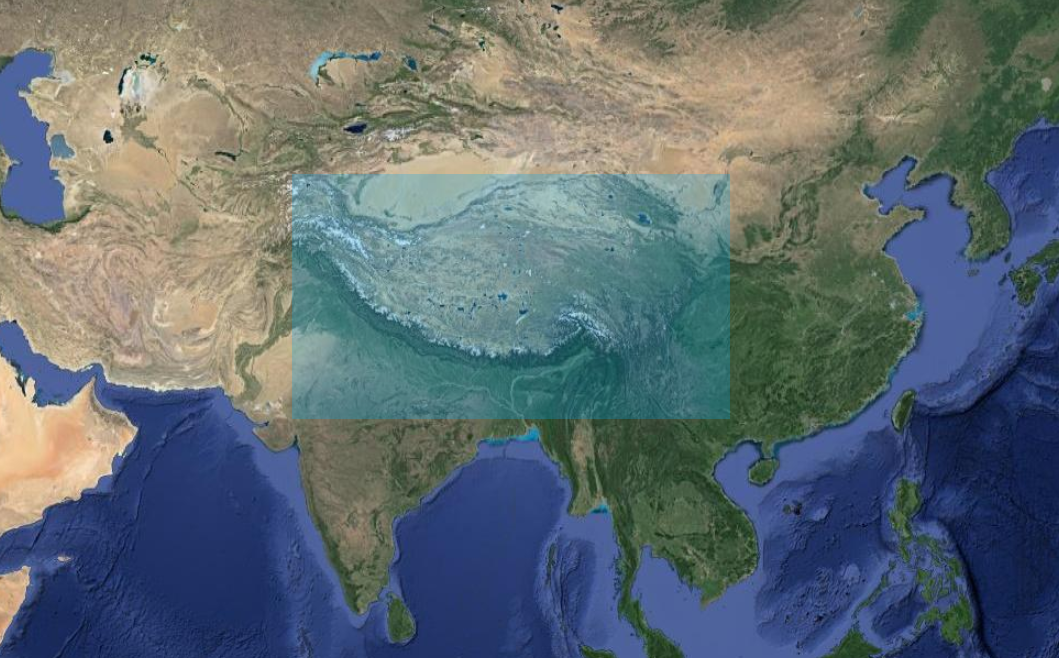}
\caption{Area used for evaluation. Latitude 24-39$^{\circ}$ N Longitude 72-107$^{\circ}$ E}
\label{fig6}
\end{figure}

To assess the performance of the algorithm and to get insight into the number of outside peaks discovered, the algorithm was applied on the surface between 24-39$^{\circ}$ N and 72-107$^{\circ}$ E. This area contains the arc of the Himalayas with the highest peaks in the world, as well as different landforms, including parts of the Tibetan Plateau, as shown in Fig.~\ref{fig6}.

The DEM used is the 3 arcsecond SRTM data collected by Jonathan de Ferranti \cite{b10}. In total, there are 576 tiles of size 1,201 x 1,201, which represents almost 2\% of the total Earth's surface.

\subsection{Results}

The algorithm was implemented in C++ and compiled using the g++ 13.2.0 compiler. The implementation is publicly available on GitHub \cite{b11}.The execution time was obtained by running the algorithm on a machine running Windows 10 Version 10.0.19045 Build 19045 with the following specifications:

\begin{itemize}
    \item Processor	Intel(R) Core(TM) i7-6700HQ CPU \@ 2.60GHz, 2601 Mhz, 4 Core(s), 8 Logical Processor(s)
    \item Installed Physical Memory RAM 16GB DDR4 2133MHz 
    \item Disk Model PM951 NVMe SAMSUNG 512GB
\end{itemize}

The time for computing the prominence values of all the mountain peaks in the region was 11 minutes and 7 seconds. Most of the time was consumed by applying the water sweep algorithm to individual tiles.

The maximum  number of outside peaks in an individual tile is 1,034 for the tile between 29-30$^{\circ}$ N, and 75-76$^{\circ}$ E. Average number of outside peaks is 403. More detailed statistics about the number of peaks are in Table~\ref{table:1}.

\begin{table}[h!]
\centering
\caption{Peak statistics}
\label{table:1}
\begin{tabular}{c | c | c | c | c }
 & \textbf{Min} & \textbf{Max} & \textbf{Median} & \textbf{Average} \\
\hline\hline
Peaks & 4,554 & 97,377 & 18,369 & 28,696.29 \\
\hline
Outside peaks & 186 & 973 & 360 & 437.13 \\
\hline
Percentage of  & 0.88\% & 6.17\%  & 1.95\% &  2.16\% \\
Outside peaks  & (445)  & (289)   & (563)  &
\end{tabular}
\end{table}

The total number of peaks in the region is 16,529,067. The initial number of outside peaks before merging was 251,790, and after merging, there were 13,333 outside peaks. So, most of the peaks have a correct prominence value found inside the analyzed surface.

\subsection{Comparison to the Kirmse and de Ferranti algorithm}

The most relevant algorithm for computing mountain prominence is the one that started from the ideas of Edward Earl and was implemented and described in a paper by Andrew Kirmse and Jonathan de Ferranti in 2017 \cite{b5}. The algorithm has two steps: creating the divide trees and then merging them. The divide trees contain all the peaks from a region composed of multiple tiles, and merging operations are not feasible for trees of that size. The authors used a smart pruning algorithm to improve complexity. The algorithm eliminates the peaks that can be proven to have a prominence value below a threshold of interest. However, the true prominence values are computed after all the divide trees are merged, so the actual prominence value of the nodes that are pruned remains unknown. This is a disadvantage compared to the algorithm presented in this paper, which eliminates the peaks for which the true prominence is known immediately and can compute the prominence value for all peaks without pruning.

A straightforward time comparison between the two algorithms is not fair because the algorithm by Andrew Kirmse and Jonathan de Ferranti is built to take advantage of multithreading and pruning and has multiple useful qualities. For example, the divide trees describe the topographic surface and the ability to compute the parent peak on the mountain ridge. However, the algorithm available  on GitHub under the MIT licence was run on the same machine on a single thread without pruning on different smaller test cases. The first part of the algorithm that computes the divide trees is, on average, slower than the whole algorithm presented in this paper. Moreover, merging divide trees without pruning becomes very expensive when the number of tiles increases.

\section{Conclusions}

This algorithm was presented as a more practical and efficient approach to computing prominence down to the dataset's precision. It is based on the observation that only a few peaks can have the peak that determines its prominence in another tile and is able to merge tiles and work only with the outside peaks and tile edges. An important further improvement would be to analyse its multithreading capabilities and adapt it to be able to process higher resolution DEM. Measurements of the surface of the Earth will continue to improve and the idea at the base of this algorithm could be further enhanced to create new and innovative algorithms for mountain prominence.


\begin{thebibliography}{00}
\bibitem{b1} Helman, Adam. The Finest Peaks-Prominence and Other Mountain Measures. Trafford Publishing, 2005
\bibitem{b2} Guth, Peter L., et al. "Digital elevation models: Terminology and definitions." Remote Sensing 13.18 (2021): 3581
\bibitem{b3} Jurgalski E (2009) Short History of Orometrical Prominence. Available at: http://www.8000ers.com/cms/en/prominence-mainmenu-179.html
\bibitem{b4} Earl Edward (2015) WinProm - Tool for calculating the topographic prominence. Available at: http://github.com/edwardearl/winprom
\bibitem{b5} Kirmse, Andrew and Jonathan de Ferranti. “Calculating the prominence and isolation of every mountain in the world.” Progress in Physical Geography 41 (2017): 788 - 802.
\bibitem{b6}Rodriguez, Ernesto, Charles S. Morris, and J. Eric Belz. "A global assessment of the SRTM performance." Photogrammetric Engineering \& Remote Sensing 72.3 (2006): 249-260
\bibitem{b7} Greg Slayden - Lists of most isolated and most prominent peaks. Available at: https://www.peakbagger.com/Default.aspx
\bibitem{b8} Cormen, Thomas H., et al. Introduction to algorithms. MIT press, 2022
\bibitem{b9} Tarjan, Robert E., and Jan Van Leeuwen. "Worst-case analysis of set union algorithms." Journal of the ACM (JACM) 31.2 (1984): 245-281
\bibitem{b10} Jonathan de Ferranti J - 1, 3 arcsecond SRTM Digital Elevation Data. Online; Available at: https://viewfinderpanoramas.org/ 
\bibitem{b11} Alex Dumitrescu (2024) Implementation of the algorithm. Avaiable at: https://github.com/Dgeorgealex/ProminenceThesis
\end{thebibliography}
\end{document}